\documentclass[10pt]{article}

\usepackage{graphicx}
\usepackage{epsfig}
\usepackage[centertags]{amsmath}
\usepackage{amssymb}

\newcommand{\rmi}{{\rm i}}

\newcommand{\la}{\langle}
\newcommand{\lla}{\left\langle}
\newcommand{\ra}{\rangle}
\newcommand{\rra}{\right\rangle}

\begin{document}

\title{A random matrix approach to decoherence}

\author{T Gorin\footnote{Thomas.Gorin@physik.uni-freiburg.de}\\
{\it\small Theoretische Quantendynamik, Fakult\" at f\" ur Physik,
           Universit\" at Freiburg, Hermann-Herder-Str. 3,}\\
{\it\small D-79104 Freiburg, Germany}\\[2mm]
T H Seligman\\
{\it\small Centro de Ciencias F\'\i sicas, University of Mexico (UNAM),
           Avenida Universidad s/n,}\\
{\it\small C.P. 62210 Cuernavaca, Morelos, Mexico}
}

\date{}

\maketitle

\begin{abstract}
In order to analyze the effect of chaos or order on the rate of
decoherence in a subsystem, we aim to distinguish effects of the
two types of dynamics by choosing initial states as random
product states from two factor spaces representing two subsystems. We
introduce a random matrix model that permits to vary the coupling
strength between the subsystems. The case of strong coupling is
analyzed in detail, and we find no significant differences except
for very low-dimensional spaces. \\


\end{abstract}


\section{Introduction}

The discussion of decoherence phenomena has been centered around 
problems of quantum optics and atomic physics, which usually implied the
use of smooth, often Gaussian, wave packets.
Yet recent ideas on how to process quantum information changed 
that situation to some extent. Indeed in this context it is desirable to 
store a maximal amount of information in product states of qubits, such that
it can be processed by unitary time evolution. During the process, the
relevant states are changed in a complicated way. Hence, they will most 
certainly resemble random states much more than Gaussian wave packets.
As the conservation of coherence is essential for any quantum information 
application, we have a clear interest in understanding decoherence 
of random states 
in such situations.
The standard tool in this context is 
the study of ``fidelity'' which involves a change in the Hamiltonian,
and a number of interesting results are available \cite{jalabert, prosen1,
prosen2}.   

The fundamental question,
 how the integrability or chaoticity of the corresponding
classical system {\it i.e.} ``quantum chaos''  affects the
process of decoherence (see {\it e.g.} \cite{ref1,ref2,ref3}), appears in
a different light, because the behaviour of random states 
becomes relevant. We shall focus on this aspect, 
concentrating our interest on the effects of the chosen dynamics. At first
glance, the semi-classical context implicit in this view may seem odd after 
the reference to quantum computing, but note that in many-body systems
some effects of the chaotic dynamics tend to survive even for very low lying 
states \cite{flores}, where semi-classical arguments certainly do not apply. 

In a recent letter \cite{gose} we proposed to discuss decoherence 
in a random matrix model, in order to separate effects of the Hamiltonian
from those of the choice of the initial pure state. Such a model is 
ideally suited to discuss the behaviour of random states. In this context 
we do not embrace the scheme of fidelity analysis \cite{prosen1},
 but simply follow decoherence as a function of time for a fixed 
Hamiltonian. 
We introduce decoherence by performing partial traces
over a subsystem instead of considering non-unitary time evolution.
What we show, could therefore more strictly be considered as a study of
the progress of entanglement of a product wave function under 
an integrable or chaotic time evolution.

In section~\ref{M} we
present our random matrix model, which allows for a variety of 
situations both with respect to the nature of the subsystems 
and the coupling strength between them. Note that models based
on random matrix theory (RMT) have been used to discuss dissipation 
(see \cite{lutz} and references therein), but these aspects will not 
be discussed here. In section~\ref{P} we shall analyze a special 
situation of strong coupling between the two subsystems.
Though not particularly relevant for quantum computation, it has the 
advantage of a transparent mathematical structure which allows to 
obtain non-perturbative analytic results. These are compared
to numerical calculations in section~\ref{N}.
As it turns out, the results are typical for other situations as well.
In particular for weak coupling, analyzed numerically 
in the second part of section~\ref{N}, we are lead to essentially the same
conclusions as in the strong coupling case.
Our approach to the problem with (largely) random wave functions 
actually separates rather nicely the effects of the dynamics from 
those of the functions considered, and in section~\ref{C} we shall 
conclude with some considerations on this point, as well as with 
a comparison of our findings with those of Refs.~\cite{prosen1, prosen2}.

\section{\label{M}The random matrix model}

Properties of chaos and integrability of a classical system manifest 
themselves both in the spectrum and in the wave functions of the 
corresponding quantum system. While the former is
invariant the latter are basis dependent. This does not mean that the
latter are irrelevant in a semi-classical context; they certainly reflect
the special features of the dynamics such as KAM tori or short periodic 
orbits. However, while localized wave packets may feel a strong influence
on their dynamics, random wave functions would be rather insensitive to
such localized features. 

To construct our RMT model we start  from the standard
quantum chaos conjecture, that
the classical ensembles \cite{cartan,ref5} ({\it e.g.} the Gaussian orthogonal
ensemble (GOE) for time reversal invariant systems) describe the universal
features of a quantum system, whose classical counterpart
is chaotic \cite{ref6}. For
classically integrable systems  we expect a random quantum spectrum 
if we exclude harmonic oscillators \cite{bertab77}. 
The random spectrum can be combined with the orthogonal invariance to give 
the so called Poisson orthogonal ensemble (POE) \cite{ref8}, extending in
this way the classical ensembles to the integrable case. This concept provides
the ideal tool to describe the evolution of random wave functions, which is
one of the main purposes of the present work.

Consider a Hamiltonian consisting of three
terms $H= H^{(0)}+ \lambda V^{(1,2)}$ with $H^{(0)} = h^{(1)}+h^{(2)}$,
where the two terms of $H^{(0)}$ act on different degrees of freedom
of the system; we may refer to them as the central system and 
the environment. Note that the $h^{(i)}$
may act on one or several degrees of freedom each, and the interaction
$V^{(1,2)}$ may or may
not induce chaos. Indeed the total system may be integrable and separable
in a different set of coordinates. 
 
Taking this Hamiltonian as a quantum operator we shall denote
by ${\cal H}_{1}$ and $ {\cal H}_{2}$ the Hilbert spaces of
the central system and the environment with dimensions $n$ and
$m$ , on which
$ h^{(1)}$ and $h^{(2)}$ respectively act. The total Hamiltonian $H$ acts
on the product space ${\cal H}={\cal H}_{1}\times {\cal H}_{2}$
 with dimension $N=nm$.
We write the basis states of ${\cal H}_{1}$ as kets with Latin letters such as
$ \vert i \rangle $ and those of ${\cal H}_{2}$ as kets with Greek letters
such as $ \vert \mu \rangle $. The states
$ \vert i, \mu \rangle = \vert i \rangle \vert \mu \rangle $
with indices conveniently written as pairs, form an eigenbasis of $H^{(0)}$.
$H$ is diagonal in a different basis, which we enumerate by a single
index $\alpha $. Thus $H_{i\mu,i^\prime \mu^\prime} = \sum_\alpha
O_{i\mu, \alpha}\; E_\alpha\; O_{i'\mu',\alpha}$
where $E_\alpha$ denotes elements of the diagonal energy matrix $E$ and $O$
the orthogonal transformation between the two bases.

We distinguish between strong and weak interaction.
The interaction strength is usually discussed in terms of the
spreading width, which indicates the width of the distribution
of the expansion coefficients of the eigenstates of $H$ in terms of those of
$H^{(0)}$. This spreading width is a semi-classical quantity
in the sense that it can be calculated from a phase space integral
\cite{BSW}. It is closely related to the width of the LDOS (the density of
the eigenstates of $H_0$ expressed in terms of those of $H$)
If the spreading width is small, {\it i.e.} if the eigenstates of
$H_0$ contain a dominant amplitude in the eigenbasis of $H$ 
we clearly have a weak coupling case, where perturbative calculations
should yield the correct answer. A detailed study will be published 
elsewhere, but we shall give some numerical results in section~\ref{N}.

We now determine appropriate matrix ensembles for the three terms in
the Hamiltonian. As mentioned before, $h^{(1)}$ could describe the 
central system and $h^{(2)}$ the environment, {\it e.g.} the heat bath. In
any case both $h^{(1)}$ and $h^{(2)}$ could pertain to either of the above
ensembles, the GOE or the POE, and $V^{(1,2)}$ could typically be 
symmetric with independent Gaussian distributed matrix elements. 

Finally we must specify the initial states. As mentioned before, we are
interested in states of complicated structure, but require that they were
initially pure. Hence we choose two random states in ${\cal H}_1$ and 
${\cal H}_2$ and define the initial state as the product state of both. The
probability measure for the random states is the orthogonally invariant one.

The situation first studied by  Zurek and coworkers \cite{ref1} 
could be simulated as a weak coupling case using a GOE with high 
level density for $h^{(2)}$ and a GOE or a POE respectively for 
$h^{(1)}$. However, we would have to use initially smooth wave
packets. In this respect the states, whose evolution is discussed
here, are entirely different.

\section{\label{P}Purity decay in the strong coupling regime}

In the case of strong coupling $h^{(1)}$ and $h^{(2)}$
determine the factor spaces ${\cal H}_{1}$ and $ {\cal H}_{2}$ only.
Their spectral properties are irrelevant except for their relative spectral
density in the energy region where the wave packet lives. The total Hamiltonian
which is essentially equal to the  interaction
$V^{(1,2)}$ will be given by the GOE for chaotic systems and by the 
POE for integrable ones. Both ensembles are
given by matrices of the form $O E O^{\tau}$, where $E$ is a diagonal energy
matrix, and $O$ is a orthogonal matrix distributed according to the Haar
measure of the orthogonal group. For the GOE the distribution of the energies
has complicated correlations and a semi circle density, while they are
independently Gaussian distributed for the POE. 

As the level density has no relation to the chaoticity 
or integrability of the system, we shall unfold both spectra to have 
uniform density with 
the variance $\la E_\alpha^2\ra$ normalized to one in the ensemble average.
As a consequence, the length of the spectrum is $2\sqrt{3}$ and the
average level spacing is $d= 2\sqrt{3}/N$. One reason to fix the 
energy scale in this way is, that the variance is easy to control
even in the weak coupling regime. 
Note that decoherence depends on the behaviour of the level density
as a function of energy. Our random matrix model allows also to choose
this behaviour differently, {\it e.g.} as dictated by the classical 
or quantum Hamiltonian.

The case of strong coupling both for integrable and chaotic systems 
was modeled in \cite{BSW} with two-dimensional anharmonic oscillators, 
and the systems considered in \cite{ref3} might be close to this domain.
We shall also use equidistant "picket fence" spectra to complete the
range of possible spectral correlations. The latter is important, because
the spectra of low dimensional systems are much stiffer \cite{ref9,berry85} 
than the universal random matrix ensembles would predict.

The dimensions $n,m$ and $N$ of our Hilbert spaces are chosen finite. 
If we think of Hamiltonians with infinite spectra this truncation is 
the only reference we make to a choice of the initial wave function.

The entanglement of the two subsystems will be measured 
in terms of the purity defined as
\cite{ref1}
\begin{equation}
I(t) = {\rm Tr}_1[{\rm Tr}_2 (\rho(t))]^2
     = {\rm Tr}_2[{\rm Tr}_1 (\rho(t))]^2 \; .
\label{equ1}\end{equation}
Here ${\rm Tr}_1$ indicates the trace with respect to the first (Latin)
index and ${\rm Tr}_2$ the one over the second (Greek) index.
The definition of the purity $I$ is related 
to the idempotency defect or linear entropy, defined as $1-I$ \cite{ref3}.

We are interested in the time-evolution of an initially pure, {\it i.e.}
non-entangled state. Therefore we construct the initial density matrix
$\rho_{i\mu, i^\prime \mu ^\prime}(0)$ from a ``product state'' which is pure 
with respect to both pairs of indices, and we have 
$I(0)= {\rm Tr}_1[{\rm Tr}_2 (\rho(0))]^2 = 1$.
Denoting by $ \Delta $ the diagonal matrix with entries
$\Delta_\alpha =\exp[\rmi t E_\alpha]$ ($\hbar$ is set equal to one), we find 
in the basis of double indices
$\rho(t  )= O\Delta O^\tau   \rho(0) O \Delta ^* O^\tau$ and by consequence
\begin{equation}
I(t  )= {\rm Tr}_1[{\rm Tr}_2(O\Delta O^\tau \rho(0) O
\Delta ^* O^\tau ) {\rm Tr}_2(O
\Delta  O^\tau    \rho(0) O
\Delta^{*} O^\tau  )] \; .
\end{equation}
We take the averages involving energies and the averages involving
states separately for different terms of the sum.
In principle we do not need to specify $\rho(0)$ in more detail. After
performing the averages, the result must be independent of the two factor
states used to construct $\rho(0)$. This is due to the orthogonal invariance
of the total Hamiltonian $H$ used, and the invariance of the purity with
respect to independent orthogonal transformations in the factor spaces.
Hence, without loss of generality we may set $\rho(0)_{11,11}=1$ with all
other matrix elements being zero. With this initial condition we obtain the
ensemble averaged purity
\begin{equation}
I(t)=\sum_{\alpha,\beta,\gamma,\delta} \,
A_{\alpha,\beta ;\gamma ,\delta} \; B_{\alpha,\beta ;\gamma,\delta}
\end{equation}
in terms of the two averages
\begin{eqnarray}
 A_{\alpha,\beta ;\gamma ,\delta} =
\langle \Delta_{\alpha} \Delta_{\gamma}
 \Delta ^{*}_{\beta} \Delta ^{*}_{\delta}\rangle
= \langle {\rm exp}[\rmi\, t(E_\alpha +E_ \gamma - E_{\beta} -E_{\delta})]
 \rangle \\
B_{\alpha,\beta ;\gamma,\delta} =\sum_{\mu,\nu,i,j}
 \lla O_{i\mu,\alpha} O_{11,\alpha}
O_{11,\beta} O_{j\mu,\beta}
O_{j\nu,\gamma} O_{11,\gamma}
O_{11,\delta} O_{i\nu,\delta}\rra \; .
\end{eqnarray}
The averages are connected only because one may force indices to be equal and
thus reduce the other to a special case; as we shall see below five different
terms exist. Note that we omitted the average symbols on the purity
itself for convenience.

The obviously relevant time scales are the Heisenberg time 
$d^{-1}= N/(2\sqrt{3})$, and the inverse length of the spectrum, which
is $1/(2\sqrt{3})$. Using these we obtain four different regimes for the 
time evolution, because of the modulus operation inherent in the 
exponential $\exp[\rmi Ht]$:
\begin{itemize}
\item[1)]{
Short times, $t\ll 1$: Here, perturbation theory can be applied, and
we will find the expected $t^2$ dependence with a factor given
to leading order by the variance of the energy eigenvalues 
$\la E_\alpha^2\ra = 1$.}
\item[2)]{
First filling of the unit circle at $t=\pi/\sqrt{3}$: We will
find a quadratic minimum for the purity with value 
$I_{\rm min}= 1/n+1/m+0(N^{-1})$.}
\item[3)]{
Long times, $2\pi/d\gg t\gg \pi/\sqrt{3}$: In this region, the spectrum has 
winded many times around the unit circle. This acts as a random number
generator eliminating correlations. Hence we will obtain a result
similar to the one for 2) though sub-leading terms may be different.}
\item[4)]{
Poincar\'e recurrence at $t= 2\pi/d$: At this point, a picket fence spectrum
will cause exact recurrence, while even for a GOE spectrum the
recurrence is essentially wiped out. Yet for low-dimensional systems with
their long-range stiffness \cite{ref9,berry85} and for models
involving harmonic oscillators this part may well be important.}
\end{itemize}
First we shall calculate the energy average $A_{\alpha,\beta;\gamma,\delta}$. 
The result does not depend on the values of the indices, but only on whether 
certain indices are equal or not. If two indices of
the energies coincide, we get either $0$ if they have opposite signs or twice
the energy if their signs are equal. It may readily be seen that five terms
are possible (the rightmost equality is valid for a random spectrum only):
\begin{equation}
\begin{split}
S_1(t) = \lla \exp[{-\rmi t \; (E_1-E_2+E_3-E_4)}]\rra \qquad &= f^4(t) \\
S_2(t) = \lla \exp[{-\rmi t \; (E_1-E_2)}]\rra &= f^2(t) \\
S_3(t) = \lla \exp[{-\rmi t \; (2E_1-E_2-E_3)}]\rra 
&= f(2t) f^2(t) \\
S_4(t) = \lla \exp[{-\rmi t \; 0}]\rra &= 1 \\
S_5(t) = \lla \exp[{-2\rmi t \; (E_1-E_2)}]\rra &= f^2(2t) \; .
\end{split}
\label{equ6}\end{equation}
The first equality in each line determines the special case at hand
and the second one gives the result for a random spectrum.
There $f(t) = \sin(\sqrt{3}\, t)/(\sqrt{3}\, t)$ is the Fourier transform 
of the level density, which we assumed to be uniform. For GOE spectra the
evaluation is more difficult, but some general considerations hold
for any kind of spectrum.
For long times all terms except $ S_{4} $ go to zero. For short times, on
the other hand, $ S_{1} $ dominates because it has the largest weight.
We now consider the four time regimes:

In the short time limit we expand the exponential.
Due to the symmetry of the energy distribution, the linear terms in
$t$ vanish while quadratic ones survive. These are of two types.
Each exponential associated with a given index has a quadratic term, and
indices in the linear terms of two exponentials may coincide. This implies
that we only need the well-known averages over monomials of fourth order
in the group elements \cite{ullahporter} to obtain
\begin{equation}
I(t) \sim 1
- 2 \langle E_{\alpha}^2 \rangle \, t^2 [1-(n+m+1)/(N+2)] \; .
\end{equation}
In the last factor we seem to have a $1/N$ correction. Yet if $n$ and $m$ grow
as $\sqrt N$ the correction is of order $1/\sqrt N$. If one of the two
dimensions is kept constant, the other becomes proportional to $N$, and
the second term is of order 1. Terms resulting from correlations
of the energies are truly of order $1/N$ and were omitted.

The next time scale is that of the first filling of the unit circle,
for which the first minimum of the function $f^2(t)$ is reached. We have a
complicated interplay of different terms and it seems that we would need the 
average over the orthogonal group $B_{\alpha,\beta;\gamma,\delta}$ completely.
However, this can be avoided using the following trick:
For uniform density of the spectrum the energy eigenvalues are essentially 
the eigenphases of a circular ensemble. For the case of GOE fluctuations 
the corresponding ensemble is known as the circular orthogonal ensemble (COE) 
\cite{ref5}, which is the ensemble of unitary symmetric $N \times N$ matrices 
$S$. This ensemble has a unique invariant measure. By identifying the 
energies with eigenphases of a COE the only approximation we make, is that 
we neglect the correlations that exist between the ends of the spectrum for 
COE. In terms of $S$ we obtain
\begin{equation}
I_{\rm min} \approx \lla {\rm Tr}_1[{\rm Tr}_2(S \,  \rho(0)\, S^*)
                       {\rm Tr}_2(S\, \rho(0)\, S^*)] \rra \; .
\end{equation}
where $S_{i\mu,j\nu} =\sum_\alpha O_{i\mu,\alpha}
{\rm exp}[\rmi E_\alpha (2\pi/\Gamma)]
O_{j\nu,\alpha}$. The ensemble average originally given as one over states and 
spectra is thus given in terms of averages over
four symmetric unitary COE matrices, two of which are complex conjugate.
Such averages are calculated in \cite{ref11} and we obtain
\begin{equation}
I_{\rm min} \approx \frac{(n+m) N^2 + \left[3(n+m)+2\right] N - 2(n+m-1)}
{N(N+1)(N+3)} \; .
\end{equation}
As we shall see below this is slightly lower than the long time limit, while
for the integrable (POE) case the long time limit and the value $I_{\rm min}$
coincide. We shall calculate the long time limit next.

For $\ t\gg 1$, the process of stretching and taking modulo $2\pi$ is a 
reasonably efficient randomizer for a fluctuating
set of numbers with correlations such as a GOE spectrum. Therefore the
eigenphases on this time scale are random both for the GOE and the POE.
Thus only the fourth term survives, where the indices of energies in conjugate
terms coincide. The energy dependence, and therefore the time dependence,
drops out and we are left with averages over the orthogonal group. Only
two-vector terms {\it i.e.} averages over elements from two rows of the matrix
occur. These have been calculated \cite{ref10}, and we find
\begin{equation}
I_\infty = \frac{(n+m)N^3 +3[4(n+m)+3]N^2+[35(n+m)+57]N + 48}
{(N+1)(N+2)(N+4)(N+6)} \; .
\end{equation}
For the POE this result holds equally at time $t=\pi/\sqrt{3}$, which we have
discussed above for the GOE, though it will oscillate for larger times.
The value for the GOE lies slightly below the one for POE. What we see is a 
weak signature of the correlation hole, characteristic of chaotic systems.

When evaluating the large $N$ limit of these expressions, we have to take 
into account the $n$ and $m$ always occur in the form $(n+m)$. 
As $N=nm$ goes to infinity, $n+m$ can behave as any power $N^q$, with 
$1/2\le q \le 1$. The two extremes are realized if $n$ and $m$ increase
simultaneously, such that the ratio $n/m$ remains constant, or if one is fixed 
such that the other becomes proportional to $N$. 
Keeping this in mind we can still expand both expressions and find
that the first minimum for the GOE case will always be slightly lower,
but the difference will diminish as $1/N^2$, while purity itself will 
either diminish as $1/\sqrt{N}$ or reach a constant value $1/n$ or $1/m$.
We can therefore conclude that the effects of spectral correlations
are quite insignificant except for the smallest dimensions in both subsystems.

For times of the order of the Heisenberg time we expect a very different
behaviour. More precisely, if  $t=2\pi/d=N\pi/\sqrt{3}$, we have exact
revival in the case of the a picket fence spectrum, while there will be no
particular signature for a random spectrum. In the case of GOE fluctuations
one should consider the width of the $k$th neighbour spacing distribution.
It is known to increase logarithmically. For $k=1$ it has a width of
$\approx 1.25 d$ and for $k=8$ its width already is $\approx 1.85 d$
\cite{french78}. 
Due to the increasing broadness, we will see no Poincar\'e revival in the
case of a GOE spectrum either. We should though note two facts: First we will 
find an additional partial revival at half the time mentioned for a picket
fence spectrum, because of the terms $S_3(t)$ and $S_5(t)$, which contain
eigen-energies with a factor two. Second, and more important, the long-range 
stiffness of spectra in low-dimensional systems \cite{ref9} implies a 
saturation of the width of the $k$th neighbour spacing distribution and could 
therefore lead to recurrence effects.

\section{\label{N}Numerical results}

\begin{figure}
\begin{center}
\includegraphics[scale=1.]{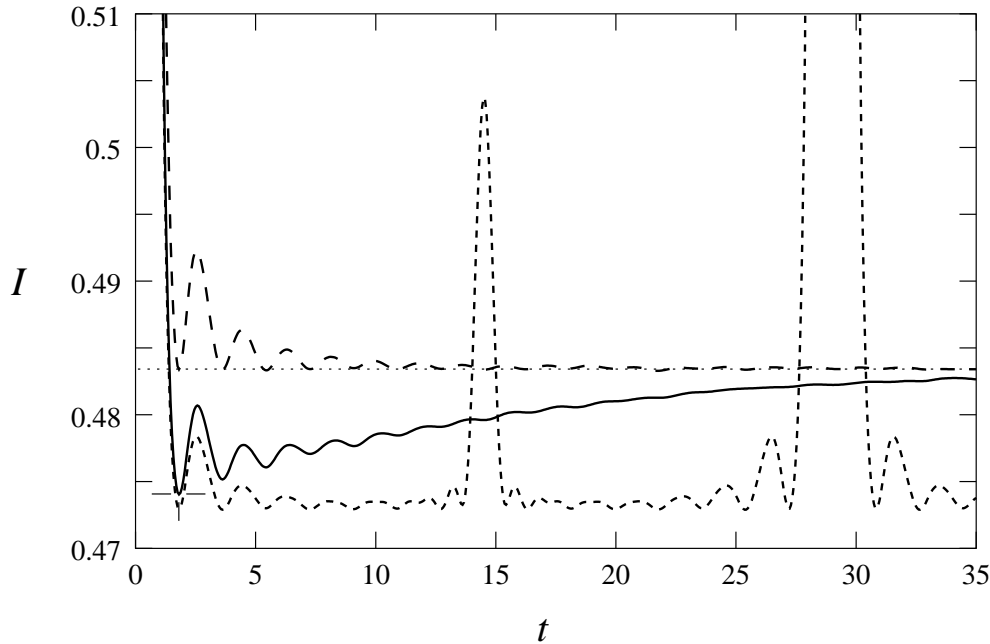}
\end{center}
\caption{The average purity as a function of time for $m=n=4$.
The ensembles considered are: POE (long dashed line), GOE (solid line), 
and picket fence (short dashed line). The value $I_\infty$ is
denoted by a thin dotted horizontal line; the value $I_{\rm min}$ at the first 
minimum by a thin upright cross. We clip a small interval of the ordinate, to 
see the difference between GOE and POE in more detail. To observe the 
behaviour of the purity at short times, see Fig.~2.}
\end{figure}

First we consider a strong coupling situation with small dimensions,
where all the effects we predict are most notable. For this purpose we choose 
the case $n=m=4$ and show the corresponding time evolution of purity in 
Fig.~1. The results are displayed for spectra with random and GOE like 
fluctuations as well as picket fence spectra. The ensemble size is $2*10^6$.  
As expected, at the beginning all curves are equal, and therefore we 
choose a scale where the quadratic dependence at the origin is not visible. 
The theoretical predictions are well fulfilled. The first minimum 
occurs at $t=\pi/\sqrt{3}$ for all spectral ensembles. 
Its depth for GOE correlations coincides with the COE result 
indicated by a cross, while both the GOE and the POE case take 
the asymptotic value given by a dotted line. In the POE case, the 
oscillations are essentially due to the form of 
$f(t)= \sin(\sqrt{3}\, t)/(\sqrt{3}\, t)$ (cf. eq.~(\ref{equ6})). 
However, the GOE and picket fence cases follow this behaviour in a similar
fashion. It is essentially a ``diffraction'' effect due 
to the sharp cutoff of the level density. {\it E. g.} for a semi-circle 
density the effect is attenuated, and for a Gaussian density 
it disappears. Therefore the experimental significance of these
oscillations is limited to special situations, 
but the rapid decay on the time scale indicated does not depend on this fact.
 The first minimum is lower for GOE like
fluctuations and for picket fence spectra than for random spectra, though
the effect is only a few percent.  The results coincide with our
theoretical predictions both at the minimum
and in the asymptotic region. This is also true for the recurrences,
which are only seen for picket fence spectra both at $t=N\pi/\sqrt{3}$ and more
weakly at $t=N\pi/(2\sqrt{3})$.

If we neglect oscillations, the rise of purity after the 
first minimum follows roughly that of
the Fourier transform of the two-point function with appropriate scaling. This
is not surprising, because we may expect that a cluster expansion of the
correlations relevant for the difference from the random case
is dominated by the two-point function. Yet it is important to 
note that the differences between the POE and GOE case lie within a
few percents for such small dimensions. If $n$ or $m$ is increased any further,
the differences become practically invisible. In \cite{gose} we show a figure 
for $n=m=10$, which confirms this fact. We conclude that decoherence of random 
states is insensitive to chaos or integrability, except for very small systems. 

\begin{figure}
\begin{center}
\includegraphics[scale=1.]{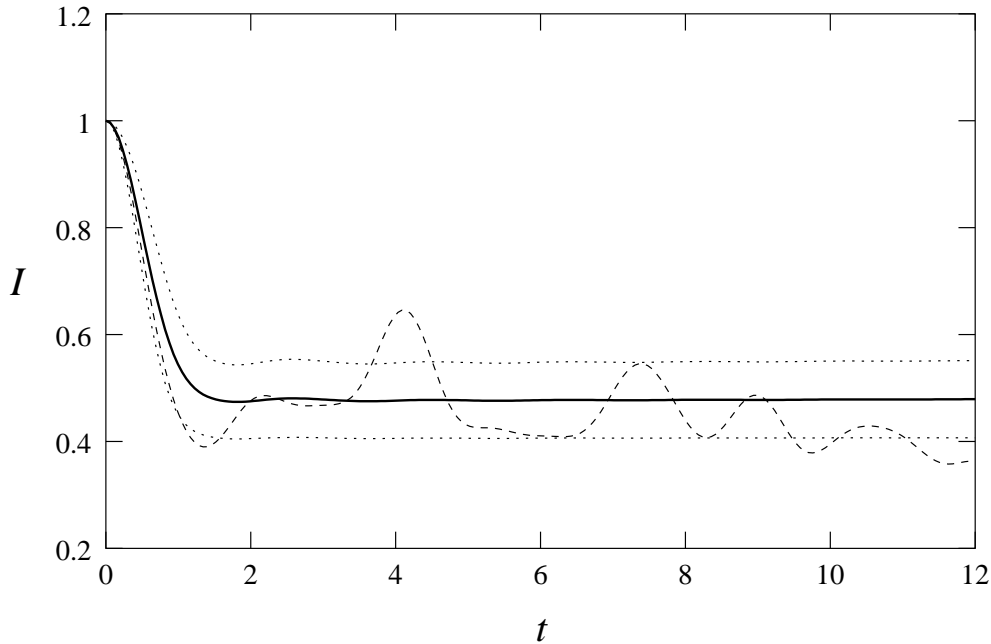}
\end{center}
\caption{The average purity as in figure 1, but for the GOE case only
(thick solid line). The average purity plus and minus the standard 
deviation (dotted lines), and the purity for the first member from the 
ensemble (dashed line).} 
\end{figure}

Concerning noticeable differences between the integrable
and chaotic cases, the news get even worse, if we study the
fluctuations of the purity as a function of time.
Fig.~2 shows the GOE case for $n=m=4$ as above but including the 
standard deviation of the purity plotted as a band around its average. 
For illustration we have also included the first member of the 
ensemble into the figure. The standard deviation  
is  big, and it is fair to say, that any medium and long
time features shown by a particular initial state in a chaotic system
are most likely fluctuations. The picture for
a random spectrum is essentially the same and thus the same conclusion 
holds for a random wave function in an integrable system. 

Calculations with larger dimensions show that the fluctuations of the purity 
diminish, but not as fast as the difference resulting from spectral
correlations. This is shown to diminish with $N^{-2}$. \\

Finally let us discuss some numerical results for weak coupling:
We construct our ensembles in the eigenbasis of $H^{(0)}$. Then $h^{(1)}$ and
$h^{(2)}$ are diagonal with elements taken from the spectrum of the GOE or the 
POE. In both cases, the spectra are unfolded to uniform density. The
matrix elements of $V^{(1,2)}$ are Gaussian distributed random variables.
As the diagonal elements are set to zero, we have:
$\la (V_{\alpha\beta}^{(1,2)})^2\ra = 1-\delta_{\alpha\beta}$. It is 
assumed, that the diagonal elements are effectively absorbed into the
Hamiltonians $h^{(1)}$ and $h^{(2)}$.

The resulting level density of the total system becomes quite complicated. 
However, it is still a simple task to obtain its variance:
\begin{equation}
\la E_\alpha^2\ra = \frac{1}{N} {\rm Tr} H^2 = \la e_i^2\ra + \la e_\mu^2\ra +
\lambda^2 (N-1)
\end{equation}
where $\la e_i^2\ra$ and $\la e_\mu^2\ra$ denote the variances of the spectra
of the subsystems, which are equal to one half. In what follows, the energy 
scale is again chosen such that $\la E_\alpha^2\ra = 1$.

While the randomness of the initial states was irrelevant in the previous 
examples of strong coupling (due to the orthogonal invariance of the total
Hamiltonian), here it makes an important difference. In distinction to 
eigenstates of $H^{(0)}$ or localized wave packets, we may expect that 
random initial states decohere faster. The reason is most easily 
understood looking at the LDOS. There a random initial state overlaps 
typically with many more eigenstates of the total system, as localized wave 
packets or eigenstates of $H^{(0)}$. In the course of time this leads to 
faster decoherence.

\begin{figure}
\begin{center}
\includegraphics[scale=1.]{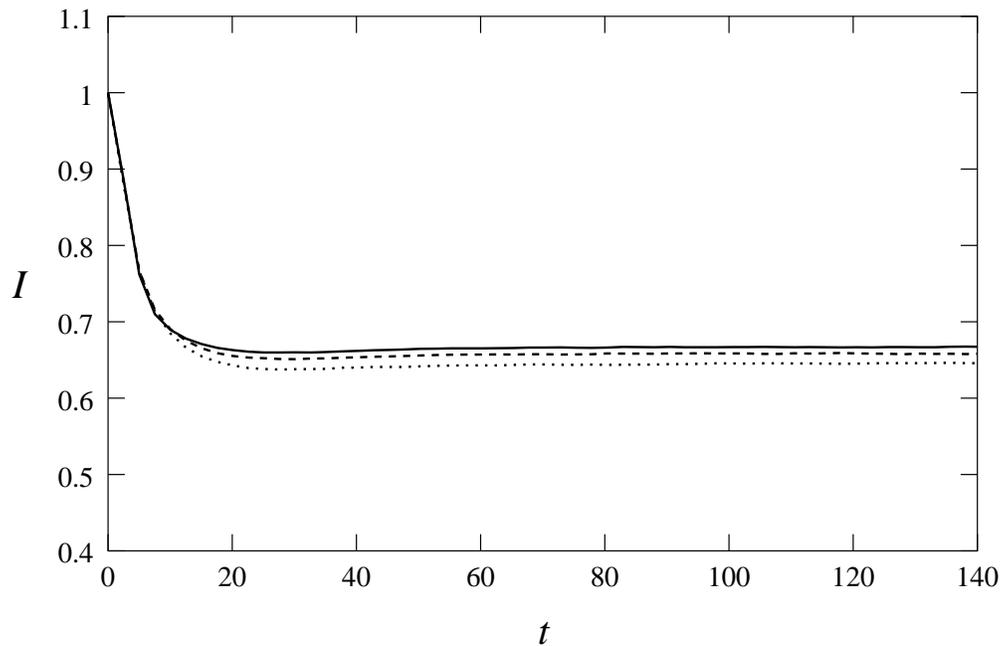}
\end{center}
\caption{The average purity as a function of time for $n=m=4$, 
in the weak coupling case ($\lambda = 0.03$). It is plotted for three 
different combinations of ensembles for $h^{(1)}$ and $h^{(2)}$: 
GOE--GOE (solid line), GOE--POE (dashed line), and POE--POE (dotted line).}
\end{figure}
 
Again small dimensions provide us with some surprises: Fig.~3 shows results 
for the dimensions $n=m=4$. The average purity is displayed for three
different combinations of spectra used in $h^{(1)}$ and $h^{(2)}$, namely 
for GOE--GOE, GOE--POE, and POE--POE. The coupling parameter is 
$\lambda = 0.03$. We find that for short and intermediate times the three 
curves have the same behaviour, but the asymptotic values differ:
In contrast to the intuitive picture, that decoherence should be strongest in
the chaotic case, the asymptotic value of the purity is lowest for the 
POE--POE case, which corresponds to two integrable subsystems. We get an
intermediate value for the GOE--POE case, and the highest value is obtained
for the GOE--GOE case, corresponding to two chaotic subsystems.

\begin{figure}
\begin{center}
\includegraphics[scale=1.]{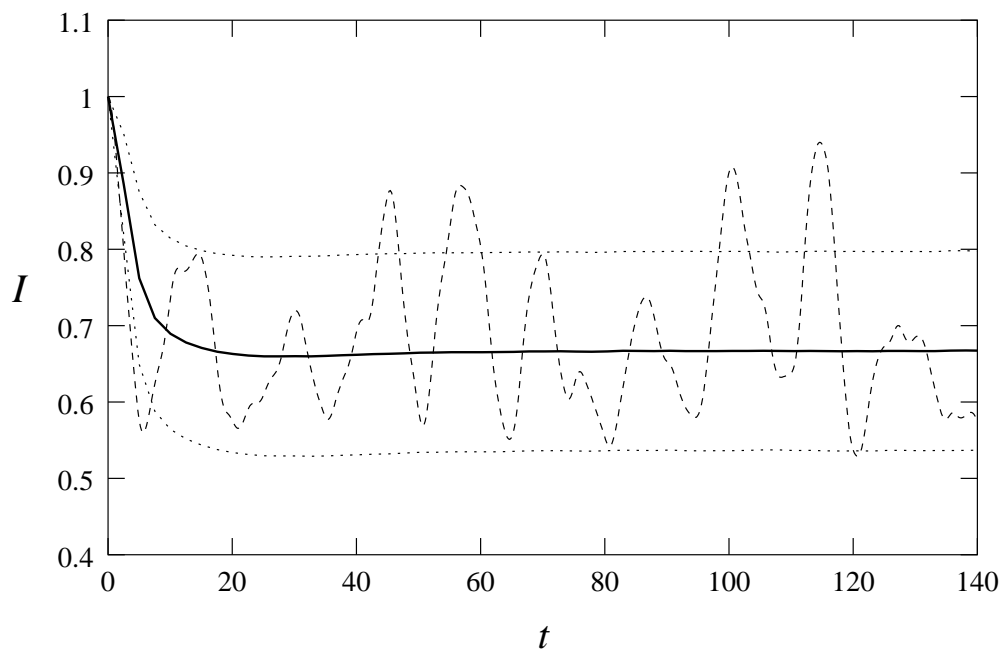}
\end{center}
\caption{The average purity as in Fig.~3, but for the GOE--GOE case only
(thick solid line). The average purity plus and minus the standard deviation 
(dotted lines), and the purity for the first member from the ensemble 
(dashed line).}
\end{figure}

Yet again the significance of this difference is limited.
This becomes clear by means of Fig.~4. There we show again the standard 
deviation of the purity in a similar fashion as in Fig.~2.
This is done for the GOE--GOE case from Fig.~3. 
In addition the purity curve for a single Hamiltonian from the ensemble is 
plotted. As can be seen in the figure, the size of the fluctuations of this 
curve is very well described by the standard deviation.

We see that the band is extremely wide and much bigger than the 
difference in the purity between the different combinations of spectral 
statistics. Note that the corresponding figure for the other two cases 
from Fig.~3 (not shown) yield the same conclusion.

\begin{figure}
\begin{center}
\includegraphics[scale=1.]{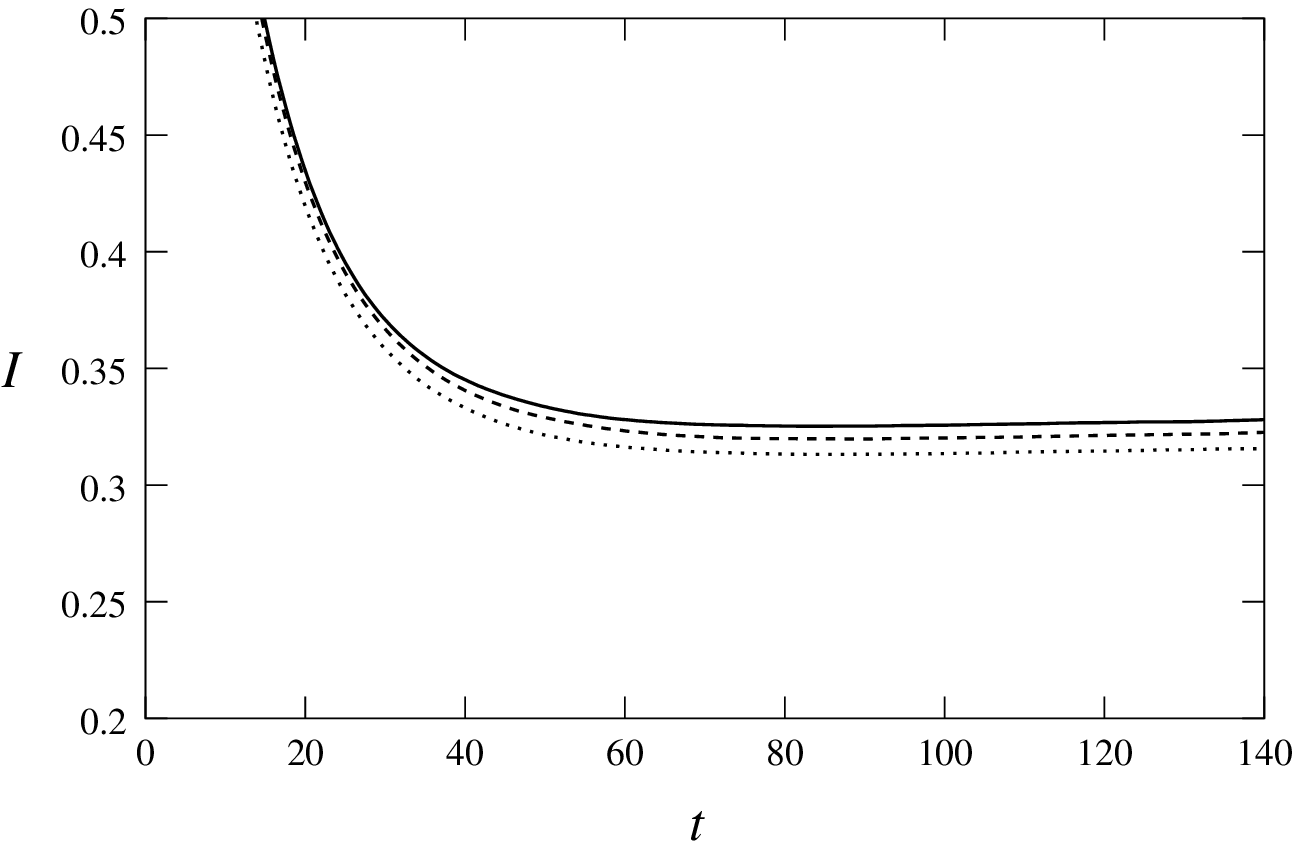}
\end{center}
\caption{The average purity as a function of time for $n=m=10$ in the
weak coupling case ($\lambda= 0.01$). It is plotted for three
different combinations of ensembles for $h^{(1)}$ and $h^{(2)}$:
GOE--GOE (solid line), GOE--POE (dashed line), and POE--POE (dotted line).}
\end{figure}

In Fig.~5 we show again the average purity for the three different combinations
of spectral statistics in the subsystems, but here the dimensions are
$n=m=10$, and the coupling parameter is $\lambda = 0.01$. The smaller coupling
parameter (as compared to the previous case, where $n=m=4$) is meant to
compensate approximately the increase of the total dimension $N$. Note that
the variance of the spectrum of $\lambda V^{(1,2)}$, which may serve as a 
measure for the strength of the perturbation, increases linearly with $N$.

The results shown in Fig.~5, are very similar to the previous ones. 
However, the differences between the three curves at large times has 
diminished. Note that an increase of $\lambda$ would further reduce these 
differences. The standard deviations of the purity curves also diminish 
(not shown), but still exceed the differences in the asymptotic average 
values. Hence, there is probably no chance to see an effect of the different
spectral statistics of the subsystems without doing averages over rather 
large samples. 
Nevertheless a detailed analysis of the different decays in weak coupling 
is an interesting open question.

\section{\label{C}Conclusions}

The influence of integrability versus chaos of the dynamics of a system on 
the entanglement of a random product wave function was considered. The type 
of dynamics was expressed in terms of spectral statistics, as this is the 
only invariant property of quantum dynamics, which is sensitive to the 
above distinction.
We conclude that the effect of the type
of dynamics on random states is small, and vanishes rapidly with the dimension
of the Hilbert spaces involved. For quantum information processing, where 
random states are typical, this means that the distinction between chaos
and integrability is not relevant to the rate of entanglement 
of two subsystems and thus for the decoherence in the central system.

Some effects have been seen for small systems, and these are interesting 
in themselves. In particular we find very large fluctuations for individual 
purity decays. One may argue that the fluctuations seen for initial packets
in the chaotic area of phase space in \cite{ref3}, could be interpreted 
as fluctuations, but clearly a large statistical calculation would 
be necessary.  

A comparison with the work of Prosen and \v Znidari\v c 
\cite{prosen1,prosen2} is more difficult. As the Hamiltonian is 
disturbed we are looking at non-linear effects even in quantum mechanics.
After all this is what the fidelity concept was designed for.
In this context our result is important because, the
reduction in the loss of fidelity,
that is obtained in  \cite{prosen1,prosen2} for chaotic dynamics, does 
not imply increased decoherence in the forward time evolution scheme with
a fixed Hamiltonian, as one might have inferred from the common concept,
that chaos enhances decoherence.

\section*{Acknowledgments}

We thank M.~C.~Nemes and H.~A.~Weidenm\"uller and T. Prosen
for stimulating discussions
and criticism as well as the A. v. Humboldt Foundation, the DGAPA (UNAM)
project IN112200  and CONACyT project 25192E for financial support.


\end{document}